\documentclass[12pt]{article}
\usepackage{amssymb,amsmath,epsfig}
\begin{document}

\title{\bf Energy Contents of a Class of Regular Black Hole Solutions in Teleparallel
Gravity}
\author{M. Sharif\thanks {msharif@math.pu.edu.pk} and Abdul
Jawad\thanks{jawadab181@yahoo.com}\\
Department of Mathematics, University of the Punjab,\\
Quaid-e-Azam Campus, Lahore-54590, Pakistan.}

\date{}

\maketitle
\begin{abstract}
In this paper, we discuss the energy-momentum problem in the realm
of teleparallel gravity. The energy-momentum distribution for a
class of regular black holes coupled with a non-linear
electrodynamics source is investigated by using Hamiltonian approach
of teleparallel theory. The generalized regular black hole contains
two specific parameters $\alpha$ and $\beta$ (a sort of dipole and
quadrupole of non-linear source) on which the energy distribution
depends. It is interesting to mention here that our results exactly
coincide with different energy-momentum prescriptions in General
Relativity.
\end{abstract}

{\bf Keywords:} Teleparallel Gravity; Energy-Momentum.\\
{\bf PACS:} 04.20.Cv; 04.20.Dw

\section{Introduction}

The notion of energy-momentum localization has been one of the
interesting, attractive but controversial issue since the advent of
General Relativity (GR). Many physicists have made painstaking
efforts on this problematic issue but it is still an open problem.
Einstein \cite{15} himself established the first energy-momentum
complex of gravitational field. After this superb work, M{\o}ller
\cite{16}, Landau-Lifshitz \cite{17}, Papapetrou \cite{18}, Bergmann
\cite{19}, Tolman \cite{20}, Weinberg \cite{21} and Komar \cite{22}
provided definitions of energy-momentum. All these complexes, except
M{\o}ller and Komar, gave reasonable results only in Cartesian
coordinates. Due to coordinate dependence and non-uniqueness,
physicists abandoned this subject for a long time.

Virbhadra \cite{23} re-aminated the subject of energy-momentum and
gave marvelous idea about coincidence of some complexes. He turned
the attention of researchers in this subject. It was found
\cite{25a}-\cite{25d} that several complexes provided the same
meaningful expressions of energy-momentum for some well-known
spacetimes. Aguireregabiria et al. \cite{26} proved that Einstein,
Landau-Lifshitz, Papapetrou, Tolman, Weinberg prescriptions yielded
the same distribution of energy, momentum and angular momentum for
any metric of Kerr-Schild class if calculations are performed in
Kerr-Schild coordinates.

Later, Virbhadra \cite{27} explored that these complexes provided
same results for a more general class than the Kerr-Schild class.
Penrose \cite{28} also shared his ideas in this field and suggested
the coordinate independent quasi-local mass. Bergqvist \cite{29}
applied seven different definitions of quasi-local mass to
Reissner-Nordstr$\ddot{o}$m and Kerr metrics and found inconsistent
results. It has been pointed out by many people
\cite{31a}-\cite{31c} that different prescriptions do not give the
same results for stringy black holes. Sharif and his co-workers
\cite{35a}-\cite{35c} found that different prescriptions did not
provide same results for some well-known spacetimes.

It was argued \cite{39,40} that telleparallel theory equivalent to
General Relativity (TEGR) might provide pointer to the localization
of energy. M{\o}ller \cite{40a} was the pioneer of tetrad theory of
gravitational field while Mikhail et al. \cite{39} derived the
energy-momentum complex in this theory. Vargas \cite{41} defined
Bergmann, Einstein and Landau-Lifshitz complexes in this alternative
theory. Sharif and Amir \cite{42} used these energy-momentum
complexes and found that results did not coincide with GR for a given
spacetime. Sharif and Nazir \cite{43} concluded that energy-momentum
turned out to be the same in TEGR and GR for Bell-Szekeres metric.

Andrade et al. \cite{44a}-\cite{44d} suggested that energy-momentum
problem might be settled down in the Hamiltonian framework of TEGR.
Maluf and his collaborators \cite{48a}-\cite{48b} accomplished
gigantic amount of work through this approach and studied energy for
some well-known spacetimes. Maluf et al. \cite{45} gave the
definitions of the gravitational energy, momentum and angular
momentum from the Hamiltonian formulation of TEGR \cite{45b}. da
Rocha-Neto and Castello-Branco \cite{49} computed gravitational
energy for the Kerr and Kerr anti-de Sitter spacetimes. Recently,
Sharif and Sumaira \cite{50} have used this procedure to obtain
energy and its relevant quantities for some vacuum and non-vacuum
spacetimes.

It has been found \cite{30}-\cite{30b} that different prescriptions
in GR yield the same results for a regular black hole and a class of
regular black holes coupled with non-linear electrodynamics source.
This paper is focussed to evaluate energy and its related
quantities for a class of regular black holes using the Hamiltonian
approach in TEGR. The layout of the paper is the following: Section
\textbf{2} contains expressions of gravitational energy, momentum,
angular momentum, gravitational and matter energy-momentum fluxes.
In section \textbf{3}, we evaluate energy and its contents for a
class of regular black hole solutions. Section \textbf{4}
provides discussion on the obtained results.

We shall use the following convention throughout the paper:
Spacetime indices $(\mu,\nu,\rho,...)$ and tangent space indices
$(a,b,c,...)$ run from 0 to 3. Time and space indices are
represented as $\mu=0,i$ and $a=(0),(i)$ respectively.

\section{Energy-Momentum in Teleparallel Theory using Hamiltonian Approach}

The basic ingredient of teleparallel theory is the tetrad field
${e^a}_\mu$ which is used to define Weitzenb\"{o}ck connection
\cite{51}
\begin{equation}\label{1}
{\Gamma^\lambda}_{\mu\nu}={e_a}^\lambda\partial_\nu{e^a}_\mu
\end{equation}
and the torsion tensor
\begin{equation}\label{2}
{T^a}_{\mu\nu}=\partial_\mu{e^a}_\nu-\partial_\nu{e^a}_\mu.
\end{equation}
The Lagrangian density for the gravitational field endowed with
matter in TEGR is described as \cite{45b}
\begin{eqnarray}\label{3}
L\equiv-\kappa e\Sigma^{abc}T_{abc}-L_M,
\end{eqnarray}
where $\kappa=1/16\pi,~e=det({e^a}_\mu)$ and the anti-symmetric
tensor $\Sigma^{abc}$ on the right two indices is
\begin{equation}\label{4}
\Sigma^{abc}=\frac{1}{4}(T^{abc}+T^{bac}-T^{cab})
+\frac{1}{2}(\eta^{ac}T^b-\eta^{ab}T^c).
\end{equation}
The corresponding field equations are
\begin{equation}\label{5}
e_{a\lambda}e_{b\mu}\partial_\nu(e\Sigma^{b\lambda\nu})
-e({\Sigma^{b\nu}}_a T_{b\nu\mu}-\frac{1}{4}e_{a\mu}
T_{bcd}\Sigma^{bcd})=\frac{1}{4\kappa}e T_{a\mu},\quad
\frac{\delta L_M}{\delta e^{a\mu}}=e T_{a\mu}.
\end{equation}
The total Hamiltonian density is \cite{52}
\begin{equation}\label{6}
H(e_{ai},\Pi_{ai})=e_{a0}C^a+\alpha_{ik}\Gamma^{ik}+\beta_k\Gamma^k
+\partial_k(e_{a0}\Pi^{ak}),
\end{equation}
where $C^a,~\Gamma^{ik},~\Gamma^k$ and $\alpha_{ik},~\beta_k$
represent primary constraints and Lagrangian multipliers
respectively.

The gravitational energy-momentum over an arbitrary volume $V$ is
defined as
\begin{equation}\label{7}
P^a=-\int_V d^3 x\partial_i\Pi^{ai},
\end{equation}
where
\begin{equation}\label{8}
-\partial_i\Pi^{ai}=\partial_i(4\kappa e\Sigma^{a0i})
\end{equation}
is the \textit{energy-momentum density} \cite{45}. The total angular
momentum can be written as \cite{53}
\begin{eqnarray}\label{9}
M^{ik}=2\kappa\int_V d^3x
e[-g^{im}g^{kj}{T^0}_{mj}+(g^{im}g^{0k}-g^{km}g^{0i}){T^j}_{mj}].
\end{eqnarray}
After some simple manipulations in the field equations (\ref{5}),
one can define the $a$ component of the \textit{gravitational
energy-momentum flux} and \textit{matter energy-momentum flux}
\cite{45a} as
\begin{equation}\label{11}
\Phi^a_g=\int_S dS_j\phi^{aj},\quad \Phi^a_m=\int_S dS_j(e{e^a}_\mu
T^{j\mu}),
\end{equation}
$S$ represents the spatial boundary of the volume $V$. Here the
quantity $\phi^{aj}$ describe the $a$ component of the gravitational
energy-momentum flux density in $j$ direction and its expression is
\begin{equation}\label{13}
\phi^{aj}=\kappa e e^{a\mu}(4\Sigma^{bcj}T_{bc\mu}
-\delta^j_\mu\Sigma^{bcd}T_{bcd}).
\end{equation}

\section{A Class of Regular Black Hole Solutions}

One of the burning issues of GR is the global regularity of black
hole solutions. Bardeen \cite{4} was the first who discovered
astonishing model known as regular black hole (also called Bardeen
model). After this, some more singularity free models
\cite{5}-\cite{7} were found also called Bardeen models \cite{8}.
These models are not exact solutions of the Einstein fields
equations due to un-availability of appropriate physical source. In
the mean time, Ayon-Beato and Garcia \cite{12}-\cite{14} found exact
singularity free solutions by coupling the EFEs with non-linear
electrodynamics. They \cite{14} generalized a regular class of exact
black hole solutions of the EFEs coupled with non-linear
electrodynamics source \cite{12}. This class of solutions can be
converted into Maxwell theory under the restriction of weak field
approximations which correspond to asymptotic
Reissner-Nordstr$\ddot{o}$m black hole. The generalized form of
these solutions is given by the line element
\begin{equation}\label{18}
ds^2=-Fdt^2+F^{-1}dr^2+r^2d\theta^2+r^2{\sin^2\theta}d\phi^2,
\end{equation}
where
\begin{equation*}
F=1-\frac{2M(r)}{r}
\end{equation*}
and the function $M(r)$ for three different models \cite{13,14} is
given by
\begin{eqnarray}\label{19}
M_1(r)&=&\frac{mr^3e^{-q^2/2mr}}{(r^2+q^2)^{3/2}},\\\label{20}
M_2(r)&=&m(1-\tanh\frac{q^2}{2mr}),\\\label{21}
M_3(r)&=&(\frac{mr^{\alpha}}{(r^2+q^2)^{\alpha/2}}
-\frac{q^2r^{\beta-1}}{2(r^2+q^2)^{\beta/2}}).
\end{eqnarray}
The associated electric field sources are given by
\begin{eqnarray}\label{23}
\Xi_1&=&\frac{qe^{-q^2/2mr}}{(r^2+q^2)^{7/2}}(r^5+\frac{(60m^2-q^2)r^4}{8m}+\frac{q^2r^3}{2}
-\frac{q^4r^2}{4m}-\frac{q^4r}{2}-\frac{q^6}{8m}),\\
\Xi_2&=&\frac{q}{4mr^3}(1-\tanh^2\frac{q^2}{2mr})(4mr-q^2\tanh\frac{q^2}{2mr}),\\
\Xi_3&=&q(\frac{{\alpha}m[5r^2-(\alpha-3)q^2]r^{\alpha-1}}{2(r^2+q^2)^{{\alpha/2}+2}}\nonumber\\
&-&\frac{[4r^4-(7{\beta}-8)q^2r^2+({\beta}-1)({\beta}-4)q^4]r^{\beta-2}}
{4(r^2+q^2)^{{\beta/2}+2}}).
\end{eqnarray}
Here $m$ and $q$ represent mass and electric
charge respectively and the parameters $\alpha,~\beta$ indicate a
sort of dipole and quadrupole moments, respectively, of non-linear
source due to the presence of asymptotic behavior of electric field.
For the choice of $\alpha\geq3,~\beta\geq4,~q\leq2s_{c}m~({s ={|
q|}/{2m}}$ and $s_{c}$ is the critical value), these solutions
elaborate regular charged black holes and geometrically its global
structure is same as Reissner-Nordstr$\ddot{o}$m black hole.
However, the disturbance occurs at essential singularity, $r=0$,
which is taken as origin of spherical coordinates. The corresponding
asymptotic behavior of these solutions is given by
\begin{eqnarray}\label{24}
F_1&=&1-\frac{2m}{r}+\frac{q^2}{r^2}+O(\frac{1}{r^3}),\nonumber\\
F_2&=&1-\frac{2m}{r}+\frac{q^2}{r^2}+O(\frac{1}{r^4}),\nonumber\\
F_3&=&1-\frac{2m}{r}+\frac{q^2}{r^2}+\alpha\frac{mq^2}{r^3}-\beta{q^4}{2r^4}
+O(\frac{1}{r^4}).
\end{eqnarray}
We can construct the tetrad components associated with (\ref{18})
by adopting the procedure \cite{42} as
\begin{equation}\label{19a}
{e^a}_\mu(r,\theta,\phi) =\left(\begin{array}{cccc}
\sqrt{F} & 0 & 0 & 0 \\
0 & \frac{1}{\sqrt{F}}\cos\phi\sin\theta & r\cos\phi\cos\theta & -r\sin\phi\sin\theta \\
0 &  \frac{1}{\sqrt{F}}\sin\phi\sin\theta & r\sin\phi\cos\theta & r\cos\phi\sin\theta \\
0 & \frac{1}{\sqrt{F}}\cos\theta & -r\sin\theta & 0\\
\end{array}
\right)
\end{equation}
with $e=det({e^a}_\mu)=r^2\sin\theta$. The non-vanishing components
of torsion tensor are
\begin{eqnarray}\label{25}
T_{(0)01}&=&\dot{\sqrt{F}},\quad
\quad T_{(1)12}=(1-\frac{1}{\sqrt{F}})\cos\theta\cos\phi,\nonumber\\
T_{(1)13}&=&-(1-\frac{1}{\sqrt{F}})\sin\theta\sin\phi,\quad
T_{(2)12}=(1-\frac{1}{\sqrt{F}})\cos\theta\sin\phi,\nonumber\\
T_{(2)13}&=&(1-\frac{1}{\sqrt{F}})\sin\theta\cos\phi,\quad
T_{(3)12}=-(1-\frac{1}{\sqrt{F}})\sin\theta
\end{eqnarray}
which yield the following non-zero components of the tensor
$T_{\lambda\mu\nu}= {e^a}_\lambda T_{a\mu\nu}$
\begin{eqnarray}\label{26}
T_{001}=\sqrt{F}\dot{\sqrt{F}},\quad
T_{212}=r(1-\frac{1}{\sqrt{F}}),\quad
T_{313}=r\sin^2\theta(1-\frac{1}{\sqrt{F}}),
\end{eqnarray}
where dot represents derivative with respect to radial component
$r$.

\subsection{Energy, Momentum and Angular Momentum}

The energy density corresponding to Eq.(\ref{18}) can be obtained
with the help of Eqs.(\ref{4}) and (\ref{8})
\begin{equation}\label{22}
-\partial_i\Pi^{(0)i}=4\kappa r\partial_1(\sin\theta(1-\sqrt{F})).
\end{equation}
Consequently, the energy will become
\begin{eqnarray}\label{27}
P^{(0)}=E&=&r[1-\sqrt{F}],\nonumber\\
E&=&r[1-\sqrt{1-2\frac{M(r)}{r}}].
\end{eqnarray}
Using the binomial expansion with $r\gg M(r)$, it follows that
\begin{equation}\label{24}
E{\approx}M(r).
\end{equation}
Inserting Eqs.(\ref{19})-(\ref{21}) in the above equation, it
follows
\begin{eqnarray}\label{29}
E_1&=&\frac{me^{-q^2/2mr}}{(1+\frac{q^2}{r^2})^{3/2}},\\\label{30}
E_2&=&m(1-\tanh(\frac{q^2}{2mr})),\\\label{28}
E_3&=&\frac{m}{(1+\frac{q^2}{r^2})^{\alpha/2}}-\frac{q^2}
{2r(1+\frac{q^2}{r^2})^{\beta/2}}.
\end{eqnarray}
Momentum and angular momentum turn out to be constant.

\subsection{Energy-Momentum Flux}

Here we evaluate energy-momentum flux. Since all the components of
gravitational energy flux density $\phi^{(0)j}$ vanish, hence the
gravitational energy flux becomes constant, i.e., for $a=0$, we have
$\Phi^{(0)}_g=\textmd{constant}$. The momentum flux density
components are
\begin{eqnarray}\label{33}
\phi^{(1)1}&=&2\kappa\sin^2\theta\cos\phi(\sqrt{F}(\sqrt{F}-1)^2),\nonumber\\
\phi^{(1)2}&=&2\kappa\sin\theta\cos\theta\cos\phi(\sqrt{F}\dot{)}(\sqrt{F}-1),\nonumber\\
\phi^{(1)3}&=&-2\kappa\sin\phi(\sqrt{F}\dot{)}(\sqrt{F}-1),\nonumber\\
\phi^{(2)1}&=&2\kappa\sin^2\theta\sin\phi(\sqrt{F}(\sqrt{F}-1)^2),\nonumber\\
\phi^{(2)2}&=&2\kappa\sin\theta\cos\theta\sin\phi(\sqrt{F}\dot{)}(\sqrt{F}-1),\nonumber\\
\phi^{(2)3}&=&2\kappa\cos\phi(\sqrt{F}\dot{)}(\sqrt{F}-1),\nonumber\\
\phi^{(3)1}&=&2\kappa\sin\theta\cos\theta(\sqrt{F}(\sqrt{F}-1)^2),\nonumber\\
\phi^{(3)2}&=&-2\kappa\sin^2\theta(\sqrt{F}\dot{)}(\sqrt{F}-1),\nonumber\\
\phi^{(3)3}&=&0.
\end{eqnarray}
The momentum flux ($\Phi^{(i)}_g$) is obtained by replacing $a=i$ in
Eq.(\ref{11})
\begin{equation}\label{34}
\Phi^{(i)}_g=\int_S dS_j\phi^{ij}.
\end{equation}
Inserting the values of momentum flux densities in the above
expression for $i=1,2,3$, we get the gravitational momentum flux
\begin{eqnarray}\label{35}
\Phi^{(1)}_g&=&-2\kappa\pi\sin\phi(\frac{F}{2}-\sqrt{F})+const,\nonumber\\
\Phi^{(2)}_g&=&2\kappa\pi\cos\phi(\frac{F}{2}-\sqrt{F})+const,\nonumber\\
\Phi^{(3)}_g&=&-4\kappa\pi\sin^2\theta(\frac{F}{2}-\sqrt{F})+const.
\end{eqnarray}
These turn out to be constant for $r\gg M(r)$, i.e.
\begin{equation*}
\frac{F}{2}-\sqrt{F}=(1/2-M(r)/r)-\sqrt{(1-2M(r)/r)}\approx-\frac{1}{2}.
\end{equation*}
giving rise to
\begin{eqnarray}\label{36}
\Phi^{(1)}_g&=&\kappa\pi\sin\phi+const,\quad
\Phi^{(2)}_g=-\kappa\pi\cos\phi+const,\nonumber\\
\Phi^{(3)}_g&=&2\kappa\pi\sin^2\theta+const.
\end{eqnarray}
Thus the components of momentum flux are free of parameters
$\alpha,~\beta,~m$ and $q$ but depend upon spherical coordinates
$\theta$ and $\phi$.

In order to evaluate matter energy-momentum flux, we have to
calculate electromagnetic energy-momentum tensor. Its non-zero
components are
\begin{eqnarray}\label{37}
T^{00}=-\frac{\Xi^{2}}{8{\pi}F},\quad
T^{11}=\frac{F\Xi^{2}}{8\pi},\quad
T^{22}=-\frac{\Xi^{2}}{8{\pi}r^2},\quad
T^{33}=-\frac{\Xi^{2}}{8{\pi}r^2\sin^2\theta}.
\end{eqnarray}
The matter energy flux becomes constant while the components of
matter momentum flux are
\begin{eqnarray}\label{38}
\Phi^{(1)}_m&=&\frac{1}{8}\sin\phi\int(r\Xi^2)dr+const,\nonumber\\
\Phi^{(2)}_m&=&-\frac{1}{8}\cos\phi\int(r\Xi^2)dr+const,\nonumber\\
\Phi^{(3)}_m&=&\frac{1}{4}\sin^{2}\theta\int(r\Xi^2)dr+const.
\end{eqnarray}
Here $\Xi$ is the electric field related to each solution and hence
matter momentum flux is different for each solution. The
non-vanishing values of $\Phi^a_g$ and $\Phi^a_m$ represent the
transfer of gravitational and matter energy-momentum respectively.

\section{Summary and Discussion}

In this paper, we have investigated gravitational energy and its
related quantities such as momentum, angular momentum, gravitational
and matter energy-momentum fluxes. These are found for a class of
regular black hole solutions of the Einstein equation coupled with a
non-linear electrodynamics source. For this purpose, we have used
the Hamiltonian approach of TEGR.

We note that Eq.(\ref{28}) gives a well-defined energy for all
values of $\alpha,~\beta,~q,~r$ except for $\alpha=0=\beta=r$. For
$\alpha=3,~\beta=4$, this yields the results of energy given in
\cite{30a}. When $\alpha=0=\beta$, the energy distribution
corresponds to the energy of Reissner-Nordstr$\ddot{o}$m spacetime.
For $q=0$, this reduces to the $\textbf{ADM}$ mass which also
corresponds to the energy of Schwarzschild solution. The energy
vanishes for $\alpha\geq 3,~\beta\geq 4,~q\leq2s_{c}m,~r=0$.

It is interesting to mention here that our results for energy
distribution (\ref{29}), (\ref{30}) are exactly the same with those
found \cite{30a,37a} by using Einstein and Bergmann prescriptions in
GR. Equation (\ref{28}) yields exactly the same energy found by Yang
et al. \cite{30b} evaluated by using Einstein and Weinberg
prescriptions in GR. The gravitational and matter energy flux vanish
and the components of momentum flux become independent of mass
parameter which turn to be constant for particular values of
$\theta$ and $\phi$. The constant gravitational momentum flux
indicates that there exist a uniform distribution of matter or no
matter in asymptotically flat region. We would like to mention here
that this prescription is coordinate independent and best tool in
the race of energy localization problem.

\end{document}